\documentclass[10pt,a4paper]{article}
\usepackage{amsmath, amssymb}
\usepackage{latexsym, color}
\usepackage{epsfig}
\usepackage{bm}

\numberwithin{equation}{section}

\newtheorem{thm}{Theorem}[section]

\newtheorem{lem}[thm]{Lemma}

\def\nm{\noalign{\medskip}}

\newcommand{\qed}{\hfill \ensuremath{\square}}
\newcommand{\ds}{\displaystyle}
\newcommand{\pf}{\noindent {\sl Proof}. \ }
\newcommand{\p}{\partial}
\newcommand{\pd}[2]{\frac {\p #1}{\p #2}}

\newcommand{\eqnref}[1]{(\ref {#1})}

\newcommand{\Rbb}{\mathbb{R}}

\def\Ba{{\bf a}}

\def\Bx{{\bf x}}
\def\By{{\bf y}}

\def\BI{{\bf I}}

\newcommand{\Ga}{\alpha}

\newcommand{\Gd}{\delta}
\newcommand{\Ge}{\epsilon}

\newcommand{\Gk}{\kappa}

\newcommand{\GU}{\Upsilon}
\newcommand{\GO}{\Omega}

\newcommand{\wBx}{\widehat{\Bx}}
\newcommand{\wBy}{\widehat{\By}}
\newcommand{\beq}{\begin{equation}}
\newcommand{\eeq}{\end{equation}}

\begin{document}
\title{Anomalous localized resonance using a folded geometry in three dimensions\thanks{\footnotesize This work was
supported by the ERC Advanced Grant Project MULTIMOD--267184, by Korean Ministry of Education, Sciences and Technology through
NRF grants No. 2010-0004091 and 2010-0017532, and by the NSF through grants DMS-0707978 and DMS-1211359.}}

\author{Habib Ammari\thanks{\footnotesize Department of Mathematics and Applications, Ecole Normale Sup\'erieure,
45 Rue d'Ulm, 75005 Paris, France (habib.ammari@ens.fr).} \and
Giulio Ciraolo\thanks{\footnotesize  Dipartimento di Matematica e
Informatica, Universit\`a di Palermo Via Archirafi 34, 90123, Palermo, Italy
  (g.ciraolo@math.unipa.it).} \and Hyeonbae
Kang\thanks{Department of Mathematics, Inha University, Incheon
402-751, Korea (hbkang@inha.ac.kr, hdlee@inha.ac.kr).}  \and
Hyundae Lee\footnotemark[4]  \and Graeme W.
Milton\thanks{\footnotesize Department of Mathematics, University
of Utah, Salt Lake City, UT 84112, USA (milton@math.utah.edu).}}

\maketitle

\begin{abstract}
If a body of dielectric material is coated by a plasmonic
structure of negative dielectric material with nonzero loss
parameter, then cloaking by anomalous localized resonance (CALR)
may occur as the loss parameter tends to zero. It was proved in \cite{acklm1, acklm2} that if the coated structure is circular (2D) and dielectric constant of the shell is a negative constant (with loss parameter), then CALR occurs, and if the coated structure is spherical (3D), then CALR does not occur. The aim of this
paper is to show that the CALR takes place if the spherical coated structure has a specially designed anisotropic dielectric tensor. The anisotropic dielectric tensor is designed by unfolding a folded geometry.
\end{abstract}

\section{Introduction}

If a body of dielectric material (core) is coated by a plasmonic structure of negative dielectric constant
with nonzero loss parameter (shell), then anomalous localized resonance may occur as the loss parameter
tends to zero. To be precise, let $\GO$ be a bounded domain in $\Rbb^d$, $d=2,3$, and $D$
be a domain whose closure is contained in $\GO$. In other words, $D$ is the core and $\GO \setminus \overline{D}$ is the shell. For a given loss parameter $\Gd>0$, the permittivity
distribution in $\Rbb^d$ is given by
 \beq
 \Ge_\Gd  = \begin{cases}
 1 \quad & \mbox{in } \Rbb^d \setminus \overline{\GO}, \\
\Ge_s+ i \Gd \quad & \mbox{in } \GO \setminus \overline{D}, \\
 \Ge_c \quad &\mbox{in } D.
 \end{cases}
 \eeq
Here $\Ge_c$ is a positive constant, but $\Ge_s$ is a negative constant representing the negative dielectric constant of the shell.
For a given function $f$ compactly supported in $\Rbb^d \setminus \overline{\GO}$ satisfying
 \beq\label{zeroint}
 \int_{\Rbb^d} f\, d\Bx=0
 \eeq
(which is required by conservation of charge), we consider the following dielectric problem:
 \beq \label{basiceqn}
 \nabla \cdot \Ge_\Gd \nabla V_\Gd = f \quad \mbox{in } \Rbb^d,
 \eeq
with the decay condition $V_\Gd (x) \to 0$ as $|x| \to \infty$. The equation \eqnref{basiceqn} is known as the quasistatic equation and the real part of $-\nabla V_\Gd (x)e^{-i\omega t}$, where $\omega$ is the frequency and $t$ is the time, represents an approximation for the physical electric field in the vicinity of $\GO$, when the wavelength of the electromagnetic radiation is large compared to $\GO$.

Let
 \beq\label{power}
 E_\Gd := \Im \int_{\Rbb^d} \Ge_\Gd |\nabla V_\Gd|^2\, d\Bx = \int_{\GO\setminus D} \Gd |\nabla V_\Gd|^2\, d\Bx
 \eeq
($\Im$ for the imaginary part),  which, within a factor proportional to the frequency, approximately represents the time averaged
electromagnetic power produced by the source dissipated into heat. Also for any region $\GU$ let
\beq E^0_\Gd(\GU)= \int_{\GU} |\nabla V_\Gd|^2\, d\Bx \eeq
which when $\GU$ is outside $\GO$ approximately represents, within a proportionality constant, the time averaged electrical energy stored in the region $\GU$. Anomalous localized resonance is the phenomenon of field
blow-up in a localized region. It may (and may not) occur
depending upon the structure and the location of the source.
Quantitatively, it is characterized by $E^0_\Gd(\GU)\to\infty$ as $\Gd\to 0$ for all regions $\GU$ that overlap the
region of anomalous resonance, and this defines that region. Cloaking due to anomalous localized resonance
(CALR) may occur when the support of the source, or part of it, lies in the anomalously resonant region. Physically the enormous
fields in the anomalously resonant region interact with the source to create a sort of optical molasses, against which the source has to do a tremendous amount of work to maintain its amplitude, and this work tends to infinity as $\Gd\to 0$. Quantitatively it is characterized by  $E_\Gd \to \infty$ as $\Gd \to \infty$.

This phenomena of anomolous resonance was first discovered by Nicorovici, McPhedran and
Milton \cite{NMM_94} and is related to invisibility cloaking
\cite{MN_PRSA_06}: the localized resonant fields created by a
source can act back on the source and mask it (assuming the source is normalized to produce
fixed power). It is also related to superlenses \cite{Pendry, Pendry2003} since, as shown in \cite{NMM_94}, the anomalous resonance
can create apparent point sources.   For these connections and further developments tied to this form of
invisibility cloaking, we refer to \cite{acklm1, acklm2,
bouchitte, bruno, MNMP} and references therein. Anomalous resonance is also presumably responsible for cloaking due to
complementary media \cite{LCZC, PR, nguyen}, although we do not study this here.

The problem of cloaking by anomalous localized resonance (CALR) can be
formulated as the problem of identifying the sources $f$ such that
 first
 \beq\label{blowup1}
E_\Gd := \int_{\GO\setminus D} \Gd |\nabla V_\Gd|^2\, d\Bx \to \infty \quad\mbox{as } \Gd \to 0,
 \eeq
and secondly, $V_\Gd/\sqrt{E_\Gd}$ goes to zero outside some
radius $a$, as $\Gd\to 0$: \beq\label{bounded} |V_\Gd
(x)/\sqrt{E_\Gd}| \to 0  \quad\mbox{as } \Gd \to 0
\quad\mbox{when}\, |x| > a. \eeq Since the quantity $E_\Gd$ is
proportional to the electromagnetic power dissipated into heat by
the time harmonic electrical field averaged over time,
\eqnref{blowup1} implies an infinite amount of energy dissipated
per unit time in the limit $\Gd\to 0$ which is unphysical. If we
rescale the source $f$ by a factor of $1/\sqrt{E_\Gd}$ then the
source will produce the same power independently of $\Gd$ and the
new associated potential $V_\Gd/\sqrt{E_\Gd}$ will, by
\eqnref{bounded}, approach zero outside the radius $a$. Hence,
cloaking due to anomalous localized resonance (CALR) occurs. The normalized source is essentially invisible from the outside, yet
the fields inside are very large.

In the recent papers \cite{acklm1, acklm2} the authors developed a
spectral approach to analyze the CALR phenomenon. In particular,
they show that if $D$ and $\GO$ are concentric disks in $\Rbb^2$
of radii $r_i$ and $r_e$, respectively, and $\Ge_s=-1$, then there
is a critical radius $r_*$ such that for any source $f$ supported
outside $r_*$ CALR does not occur, and for sources $f$ satisfying
a mild (gap) condition CALR takes place. The critical radius $r_*$
is given by $r_* = \sqrt{r_e^3/r_i}$ if $\Ge_c=1$, and by $r_*=
r_e^2/r_i$ if $\Ge_c \neq 1$. It is also proved that if $\Ge_s
\neq -1$, then CALR does not occur: $E_\Gd$ is bounded regardless
of $\Gd$ and the location of the source. It is worth mentioning
that these results (when $\Ge_c=-\Ge_s=1$) were extended in
\cite{klsw} to the case when the core $D$ is not radial by a
different method based on a variational approach. There the source $f$ is assumed to be supported on circles. 

The situation in three dimensions is completely different. If $D$
and $\GO$ are concentric balls in $\Rbb^3$, CALR does not occur
whatever $\Ge_s$ and $\Ge_c$ are, as long as they are constants.
We emphasize that this discrepancy comes from the convergence rate
of the singular values of the Neumann-Poincar\'e-type operator
associated with the structure. In 2D, they converge to $0$
exponentially fast, but in 3D they converge only at the rate of
$1/n$. See \cite{acklm2}. The absence of CALR in such coated
sphere geometries is also linked with the absence of perfect plasmon waves:
see the appendix in \cite{klsw}. On the other hand, in a slab geometry
CALR is known to occur in three dimensions with a single dipolar source
\cite{MN_PRSA_06}. (CALR is also known to occur for the full time-harmonic Maxwell equations with a single
dipolar source outside the slab superlens \cite{dong, MN_PRSA_06, Xiao}.)

\begin{figure}[h!]
\begin{center}
\epsfig{figure=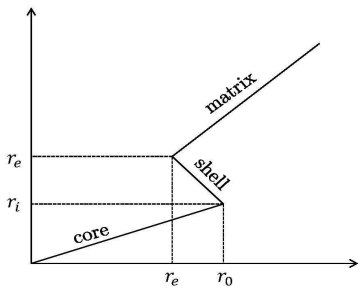, height=4cm, width=4.8cm}
\end{center}
\caption{unfolding map} \label{fig1}
\end{figure}

The purpose of this paper is to show that we are able to make CALR
occur in three dimensions by using a shell with a specially
designed anisotropic dielectric constant. In fact, let $D$ and
$\GO$ be concentric balls in $\Rbb^3$ of radii $r_i$ and $r_e$,
and choose $r_0$ so that $r_0>r_e$. For a given loss parameter
$\Gd>0$, define the dielectric constant $\bm{\Ge}_\Gd$ by
\beq\label{bmge} \bm{\Ge}_\Gd (\Bx)=
\begin{cases}
\BI,\quad & |\Bx|>r_e,\\
\ds (\Ge_s + i \Gd) a^{-1} \left(\BI+ \frac{b(b-2|\Bx|)}{|\Bx|^2} \widehat{\Bx} \otimes \widehat{\Bx} \right), \quad & r_i< |\Bx|<r_e,\\
\ds \Ge_c \sqrt{\frac{r_0}{r_i}} \BI, \quad & |\Bx|<r_i,
\end{cases}
\eeq where $\BI$ is the $3 \times 3$ identity matrix, $\Ge_s$ and
$\Ge_c$ constants, $\widehat{\Bx}=\frac{\Bx}{|\Bx|}$, and
\beq\label{ab} a:=\frac{r_e - r_i}{r_0 -r_e}>0, \quad b:=(1+a)r_e.
\eeq Note that $\bm{\Ge}_\Gd$ is anisotropic and variable in the
shell. This dielectric constant is obtained by push-forwarding
(unfolding) that of a folded geometry as in Figure \ref{fig1}.
(See the next section for details.) It is worth mentioning that
this idea of a folded geometry has been used in \cite{MNMCJ_09}
to prove CALR in the analogous two-dimensional cylinder structure for
a finite set of dipolar sources. Folded geometries were first
introduced in \cite{leonhardt} to explain the properties of superlenses,
and their unfolding map was generalized in  \cite{MNMCJ_09} to allow
for three different fields, rather than a single one, in the overlapping regions.
Folded cylinder structures were studied as superlenses in \cite{Yan} and folded geometries using
bipolar coordinates were introduced in \cite{chenchan} to obtain new complementary media cloaking structures. More general
folded geometries were rigorously investigated in \cite{nguyen}.

For a given source $f$ supported outside $\overline{B_{r_e}}$ let $V_\Gd$ be the solution to
\beq\label{dieqn}
\begin{cases}
\nabla\cdot(\bm{\Ge}_\Gd \nabla  V_\Gd) =f\quad \mbox{in}~ \mathbb{R}^3,\\
V_\Gd (\Bx)\rightarrow 0 \quad \mbox{as}~ |\Bx| \rightarrow \infty,
\end{cases}
\eeq and define \beq\label{conserv} E_\Gd=\Im \int_{\Rbb^3}
\bm{\Ge}_\Gd \nabla V_\Gd \cdot\nabla \overline{V_\Gd}\, d \Bx,
\eeq where $\overline{V_\Gd}$ is the complex conjugate of $V_\Gd$.
Let $F$ be the Newtonian potential of the source $f$, {\it i.e.},
\beq F(\Bx):= \int_{\Rbb^3}G(\Bx-\By) f(\By) d \By, \eeq with
$G(\Bx-\By)= -\frac{1}{4\pi |\Bx-\By|}$. Since $f$ is supported in
$\Rbb^3 \setminus \overline{B_{r_e}}$, $F$ is harmonic in $|\Bx| <
R$ for some $R>r_e$ and can be expressed there as \beq F(\Bx)=
\sum_{n=0}^\infty\sum_{k=-n}^n f_{n}^{k} |\Bx|^n Y_n^k(\wBx), \eeq
where $Y_n^k(\wBx)$ is the (real) spherical harmonic of degree
$n$ and order $k$.

The following is the main result of this paper.

\begin{thm}\label{mainthm}
\begin{itemize}
\item[(i)]  If $\Ge_c = -\Ge_s = 1$, then weak CALR occurs and the
critical radius is $r_*=\sqrt{ r_er_0}$, {\it i.e.}, if the source
function $f$ is supported inside the sphere of radius $r_*$ (and
its Newtonian potential does not extend harmonically to $\Rbb^3$),
then \beq \limsup_{\Gd \to 0} E_\Gd = \infty, \eeq and there
exists a constant $C$ such that \beq\label{bounded1}
|V_\delta(\Bx)|<C \eeq for all $\Bx$ with
$|\Bx|>{r_0^{2}}{r_e}^{-1}$. If, in addition, the Fourier
coefficients $f_n^k$ of $F$ satisfy the following gap condition:
\begin{quote}
{\rm [GC1]}: There exists  a sequence $ \{n_j\} $ with
$n_1<n_2<\cdots$ such that
$$
\lim_{j \to \infty} \rho^{n_{j+1}-n_j} \sum_{k=-n_j}^{n_j}  {n_j}r_*^{2{n_j}}|f_{n_j}^{k}|^2 = \infty
$$
\end{quote}
where $\rho:=r_e/r_0$, then CALR occurs, {\it i.e.},
\beq
\lim_{\Gd \to 0} E_\Gd =\infty,
\eeq
and $V_\Gd/\sqrt{E_\Gd}$ goes to zero outside the radius ${r_0^{2}}/{r_e}$, as implied by \eqnref{bounded1}.

\item[(ii)]  If $\Ge_c \neq -\Ge_s = 1$, then weak CALR occurs and
the critical radius is $r_{**}=r_0$. If, in addition, the Fourier coefficients $f_n^k$ of $F$ satisfy
\begin{quote}
{\rm [GC2]}: There exists  a sequence $ \{n_j\} $ with
$n_1<n_2<\cdots$ such that
$$
\lim_{j \to \infty} \rho^{2(n_{j+1}-n_j)} \sum_{k=-n_j}^{n_j}  {n_j}r_0^{2{n_j}}|f_{n_j}^{k}|^2 = \infty,
$$
\end{quote}
then CALR occurs.

\item[(iii)]  If $-\Ge_s \neq 1$, then CALR does not occur.
\end{itemize}
\end{thm}

We emphasize that [GC1] and [GC2] are mild conditions on the
Fourier coefficients of the Newtonian potential of the source
function. For example, if the source function is a dipole in
$B_{r_*} \setminus \overline{B}_e$, {\it i.e.}, $f(\Bx)=\Ba
\cdot\nabla\delta_\By(\Bx)$ for a vector $\Ba$ and $\By\in
B_{r_*}\setminus \overline{B}_e$ where $\delta_\By$ is the Dirac
delta function at $\By$, [GC1] and [GC2] hold and CALR takes
place. A proof of this fact is provided in the appendix. Similarly
one can show that if $f$ is a quadrupole,
$f(x)=A:\nabla\nabla\delta_\By(\Bx) = \sum_{i,j=1}^2 a_{ij}
\frac{\p^2}{\p x_i \p x_j} \delta_\By(\Bx)$ for a $3 \times 3$
matrix $A=(a_{ij})$ and $\By\in B_{r_*}\setminus \overline{B}_e$,
then [GC1] and [GC2] hold.

\section{Proof of Theorem \ref{mainthm}}

Let $r_i$, $r_e$ and $r_0$ be positive constants satisfying $r_i <
r_e < r_0$, as before. For given constants $\Gk_c$, $\Gk_s$ and
$\Gk_m$, and a source $f$ supported outside $\overline{B_{r_e}}$,
let $u_c$, $u_s$ and $u_m$ be the functions satisfying \beq
\label{main_eq_folded}
\begin{cases}
\triangle u_c =0\quad \mbox{in}~ B_{r_0},\\
\triangle u_s =0\quad \mbox{in}~ B_{r_0}\setminus \overline{B_{r_e}},\\
\triangle u_m =f\quad \mbox{in}~ \mathbb{R}^3\setminus \overline{B_{r_e}},\\
\ds u_c = u_s,\quad \Gk_c \pd{u_c}{r} = \Gk_s \pd{u_s}{r} \quad \mbox{on}~ \partial B_{r_0},\\
\nm
\ds u_s = u_m,\quad \Gk_s \pd{u_s}{r} = \Gk_m \pd{u_m}{r} \quad \mbox{on}~ \partial B_{r_e},\\
u_m(\Bx)\rightarrow 0 \quad \mbox{as}~ |\Bx|\rightarrow \infty.
\end{cases}
\eeq We emphasize that the domains of $u_c$, $u_s$, and $u_m$ are
overlapping on $r_e \le |\Bx| \le r_0$ so that the solutions
combined may be considered as the solution of the transmission
problem with dielectric constants $\Gk_c$, $\Gk_s$ and $\Gk_m$ in
the folded geometry as shown in Figure \ref{fig1}. We unfold the
solution into one whose domain is not overlapping, following the
idea in \cite{MNMCJ_09}.

In terms of spherical coordinates $(r, \theta, \phi)$, the unfolding map $\Phi=\{ \Phi_c, \Phi_s, \Phi_m \}$ is given by
\beq
\begin{cases}
\Phi_m (r, \theta, \phi) =  (r, \theta, \phi), \quad & r \ge r_e,\\
\ds \Phi_s (r, \theta, \phi) = \left( r_e- \frac{r_e - r_i}{r_0 - r_e}(r-r_e), \theta, \phi \right), \quad & r_e \le r \le r_0,\\
\ds \Phi_c (r, \theta, \phi) = \left( \frac{r_i}{r_0}r, \theta, \phi \right), & r \le r_0.
\end{cases}
\eeq
Then the folding map can be written (with an abuse of notation) as
\beq
\Phi^{-1}(\Bx)=
\begin{cases}
\ds \Bx,\quad & \quad |\Bx|>r_e,\\
\ds - a \Bx + b \widehat{\Bx},& \quad r_i< |\Bx|<r_e,\\
\ds \frac{r_0}{r_i}\Bx, & \quad |\Bx|<r_i,
\end{cases}
\eeq where $a$ and $b$ are constants defined in \eqnref{ab}.

Let $\bm{\Gk}(\Bx)$ be the push-forward by the unfolding map $\Phi$, namely,
\beq \label{kappa_push}
\bm{\Gk}(\Bx)=
\begin{cases}
\Gk_m |\det \nabla \Phi_m(\By)|^{-1} \nabla \Phi_m(\By) \nabla \Phi_m(\By)^t ,\quad & |\Bx|>r_e,\\
- \Gk_s |\det \nabla \Phi_s(\By)|^{-1} \nabla \Phi_s(\By) \nabla \Phi_s(\By)^t, \quad & r_i< |\Bx|<r_e,\\
\Gk_c |\det \nabla \Phi_c(\By)|^{-1} \nabla \Phi_c(\By) \nabla \Phi_c(\By)^t, \quad & |\Bx|<r_i,
\end{cases}
\eeq
where $\Bx=\Phi(\By)$. By straight-forward computations one can see that $\bm{\Gk}=\bm{\Ge}$ given in \eqnref{bmge} if we set
\beq\label{GkGe}
\Gk_m=1, \quad \Gk_s=-(\Ge_s+i\Gd), \quad \Gk_c=\Ge_c.
\eeq
Moreover, the solution $V_\Gd$ to \eqnref{dieqn} is given by
\beq
V_\Gd (\Bx)=
\begin{cases}
u_m\circ \Phi^{-1} (\Bx) \quad &\mbox{if}~ |\Bx|>r_e,\\
u_s\circ \Phi^{-1} (\Bx) \quad &\mbox{if}~ r_i< |\Bx|<r_e,\\
u_c\circ \Phi^{-1} (\Bx) \quad &\mbox{if}~ |\Bx|<r_i,
\end{cases}
\eeq and by the change of variables $\Bx = \Phi_s(\By)$ and
\eqnref{kappa_push}, we have \beq\label{energy2} E_\Gd= \Im
\int_{\Rbb^3} \bm{\Ge}(\Bx)  \nabla  V_\Gd (\Bx)\cdot \nabla
\overline{V_\Gd(\Bx)} = \Gd \int_{r_e < |\By| < r_0} \left| \nabla
u_s(\By) \right|^2 . \eeq

Suppose that the source $f$ is supported in $|\Bx| >R$ for some
$R>r_e$. Then the solution $u$ to \eqnref{main_eq_folded} can be
expressed in $|\Bx| <R$ as follows: \beq
\begin{cases}
u_c (\Bx) =\ds\sum_{n=0}^\infty\sum_{k=-n}^n a_n^k |\Bx|^n Y_n^k (\wBx), \quad &\mbox{if}~ |\Bx| <r_0,\\
u_s(\Bx) =\ds\sum_{n=0}^\infty\sum_{k=-n}^n (b_n^k |\Bx|^n + c_n^k |\Bx|^{-n-1}) Y_n^k (\wBx), \quad &\mbox{if}~ r_e< |\Bx| <r_0,\\
u_m(\Bx) =\ds\sum_{n=0}^\infty\sum_{k=-n}^n(e_n^k |\Bx|^n + d_n^k |\Bx|^{-n-1}) Y_n^k (\wBx), \quad &\mbox{if}~ r_e< |\Bx| <R,
\end{cases}
\eeq where the coefficients satisfy the following relations
resulting from the interface conditions:
\begin{align*}
a_n^k r_0^n &= b_n^k r_0^n + c_n^k r_0^{-n-1},\\
e_n^k r_e^n +d_n^k r_e^{-n-1}&= b_n^k r_e^n + c_n^k r_e^{-n-1},\\
\Gk_c a_n^k n r_0^n& = \Gk_s ( b_n^k n r_0^n -c_n^k (n+1)r_0^{-n-1}),\\
\Gk_s(b_n^k nr_e^n - c_n^k (n+1) r_e^{-n-1})& = \Gk_m ( e_n^k n r_e^n -d_n^k (n+1) r_e^{-n-1}).
\end{align*}
By solving this system of linear equations one can see that
$$
a_n^k= a_n e_n^k, \quad b_n^k= b_n e_n^k, \quad c_n^k= c_n e_n^k, \quad d_n^k= d_n e_n^k,
$$
where
\begin{align}
a_n=\frac{-\rho^{2n+1}(2n+1)^2\Gk_m\Gk_s }{(n^2+n)(\Gk_s -\Gk_c)(\Gk_s - \Gk_m) - \rho^{2n+1}((n+1)\Gk_s+n\Gk_c)((n+1)\Gk_m +n\Gk_s)},\\
b_n=\frac{-\rho^{2n+1}\Gk_m(2n+1)((n+1)\Gk_s + n \Gk_c)}{(n^2+n)(\Gk_s -\Gk_c)(\Gk_s - \Gk_m) - \rho^{2n+1}((n+1)\Gk_s+n\Gk_c)((n+1)\Gk_m +n\Gk_s)},\\
c_n=\frac{-r_e^{2n+1}\Gk_m n(2n+1)(\Gk_s-\Gk_c)}{(n^2+n)(\Gk_s -\Gk_c)(\Gk_s - \Gk_m) - \rho^{2n+1}((n+1)\Gk_s+n\Gk_c)((n+1)\Gk_m +n\Gk_s)},\\
d_n=-\frac{nr_e^{2n+1}[\rho^{2n+1}(\Gk_m-\Gk_s)((n+1)\Gk_s+n\Gk_c) + (\Gk_s-\Gk_c)(n\Gk_m+(n+1)\Gk_s)]}{(n^2+n)(\Gk_s -\Gk_c)(\Gk_s - \Gk_m) - \rho^{2n+1}((n+1)\Gk_s+n\Gk_c)((n+1)\Gk_m +n\Gk_s)}.
\end{align}
Here $\rho$ is defined to be $r_e/r_0$

Let $F$ be the Newtonian potential of $f$, as before. Since $u-F$
is harmonic in $|\Bx| > r_e$ and tends to $0$ as $|\Bx| \to
\infty$, we have \beq e_n^k=f_n^k. \eeq So $u_m$ (the solution in
the matrix) is given by \beq u_m (\Bx)= F(\Bx) +
\sum_{n=0}^\infty\sum_{k=-n}^n f_{n}^{k}d_n |\Bx|^{-n-1}
Y_n^k(\wBx). \eeq Since $|d_n| \le C r_0^{2n}$, we have \beq
|u_m(\Bx) -F (\Bx)|\le C\sum_{n=0}^\infty\sum_{k=-n}^n
|f_{n}^{k}|r_0^{2n} |\Bx|^{-n-1}<\infty \eeq if
$|\Bx|=r>{r_0^{2}}{r_e}^{-1}$. This proves \eqnref{bounded1}.

The solution in the shell $u_s$ is given by \beq u_s (\By)=
\sum_{n=0}^\infty\sum_{k=-n}^n f_{n}^{k} (b_n |\By|^n + c_n
|\By|^{-n-1}) Y_n^k(\wBy). \eeq Using Green's identity and the
orthogonality of $Y_n^k$, we obtain that
\begin{align*}
&\int_{r_e < |\By| < r_0} \left| \nabla  u_s (\By) \right|^2 = \int_{|\By|=r_0}   u_s \overline{\pd{u_s}{r}}- \int_{|\By|=r_e}   u_s \overline{\pd{u_s}{r}}\\
&=\sum_{n=0}^\infty\sum_{k=-n}^n |f_{n}^{k}|^2\left[ (b_n r_0^n + c_n r_0^{-n-1}) (n\overline{b_n} r_0^{n} -(n+1)\overline{c_n} r_0^{-n-1})r_0\right]\\
&\quad -\sum_{n=0}^\infty\sum_{k=-n}^n |f_{n}^{k}|^2\left[(b_n r_e^n + c_n r_e^{-n-1}) (n\overline{b_n} r_e^{n} -(n+1)\overline{c_n} r_e^{-n-1})r_e\right]\\
&=\sum_{n=0}^\infty\sum_{k=-n}^n |f_{n}^{k}|^2\left[n|b_n|^2(r_0^{2n+1} -r_e^{2n+1}) -(n+1)|c_n|^2 (r_0^{-2n-1} -r_e^{-2n-1})\right]\\
&\approx\sum_{n=0}^\infty\sum_{k=-n}^n n|f_{n}^{k}|^2\left(|b_n|^2 r_0^{2n+1} +|c_n|^2 r_e^{-2n-1}\right) .
\end{align*}
The estimate \eqnref{energy2} yields \beq\label{energy3} E_\Gd \approx
\Gd \sum_{n=0}^\infty\sum_{k=-n}^n n|f_{n}^{k}|^2\left(|b_n|^2
r_0^{2n+1} +|c_n|^2 r_e^{-2n-1}\right). \eeq

\medskip
\noindent (i) Suppose that $\Ge_c = -\Ge_s =1$. With the notation in \eqnref{GkGe}, we have
\begin{align*}
|(n^2+n)(\Gk_s -\Gk_c)(\Gk_s - \Gk_m) - \rho^{2n+1}((n+1)\Gk_s+n\Gk_c)((n+1)\Gk_m +n\Gk_s)| \approx n^2 (\Gd^2 + \rho^{2n+1}),
\end{align*}
and hence
\beq
|b_n| \approx \frac{\rho^{2n}}{\delta^2 + \rho^{2n}},\quad |c_n| \approx  \frac{ \delta r_e^{2n}}{ \delta^2 + \rho^{2n}}.
\eeq
It then follows from \eqnref{energy3} that
\beq
E_\Gd \approx\sum_{n=0}^\infty\sum_{k=-n}^n \frac{\Gd nr_e^{2n}|f_{n}^{k}|^2}{\delta^2+\rho^{2n}}.
\eeq

Let
 \beq
 N_\delta = \frac{\log \delta}{\log \rho}.
 \eeq
If $n \le N_\delta$, then we have that $\delta \le \rho^{|n|}$, and hence
\begin{align*}
\sum_{n=0}^\infty\sum_{k=-n}^n \frac{\Gd nr_e^{2n}|f_{n}^{k}|^2}{\delta^2+\rho^{2n}}
& \ge  \sum_{n \le N_\delta}\sum_{k=-n}^n \frac{\Gd nr_e^{2n}|f_{n}^{k}|^2}{\delta^2+\rho^{2n}} \\
& \ge   \Gd mr_0^{2m}\sum_{k=-m}^{m}|f_{m}^{k}|^2\ge  \frac{\Gd m}{2m+1}r_0^{2m}\left(\sum_{k=-m}^{m}|f_{m}^{k}|\right)^2
\end{align*}
for any $m\le N_\delta$. By taking $\delta$ to be $\rho^n$, $n=1,2,\ldots$, we see that if the following holds
\beq\label{limsup-cond}
\limsup_{n\rightarrow\infty}{ (r_er_0)^{n/2}\sum_{k=-n}^n|f_{n}^{k}|} =
\infty,
\eeq
then there is a sequence $\{n_k\}$ such that
\beq \lim_{k \to \infty}
E_{\rho^{|n_k|}} = \infty,
\eeq
{\it i.e.},  weak CALR occurs.

Suppose that the source function $f$ is supported inside the
critical radius $r_*=\sqrt{ r_er_0}$ (and outside $r_e$) and its
Newtonian potential $F$ cannot be extended harmonically in $|x| <
r_*$. Then
 \beq\label{fser}
 \limsup_{n\rightarrow\infty}\left(\sum_{k=-n}^n |f_n^k|\right)^{1/n}> 1/\sqrt{r_er_0}.
 \eeq
and consequently, \eqnref{limsup-cond} holds. This proves that if
the source function $f$ is supported inside the sphere of critical
radius $r_*$, then weak CALR occurs.

If the source function $f$ is supported outside the sphere of
critical radius $r_*=\sqrt{ r_er_0}$, then its Newtonian potential
$F$ can be extended harmonically in $|x| < r_*+2\eta$ for $\eta>0$
and \beq \sum_{n=0}^\infty\sum_{k=-n}^n \frac{\Gd
nr_e^{2n}|f_{n}^{k}|^2}{\delta^2+\rho^{2n}}  \le
\sum_{n=0}^\infty\sum_{k=-n}^n nr_*^{2n}|f_{n}^{k}|^2  \le C \|
F\|_{L^2(\partial B_{r_*+\eta})}^2< \infty.
 \eeq
So $E_\Gd$ is bounded regardless of $\Gd$ and CALR does not occur.

Suppose that $f$ is supported inside $r_*$ and [GC1] holds. Let $ \{n_j\} $ be the sequence satisfying
$$
\lim_{j \to \infty} \rho^{n_{j+1}-n_j} \sum_{k=-n_j}^{n_j}  {n_j}r_*^{2{n_j}}|f_{n_j}^{k}|^2 = \infty.
$$
If $\delta=\rho^{\alpha}$ for some $\Ga$, let $j(\alpha)$ be the number in the sequence such that
$$
n_{j(\alpha)} \le \alpha < n_{j(\alpha)+1}.
$$
Then, we have
 \begin{align*}
 E_\Gd \approx \sum_{n \le N_\delta}\sum_{k=-n}^n \frac{\Gd nr_e^{2n}|f_{n}^{k}|^2}{\delta^2+\rho^{2n}}  &\ge \rho^{\alpha}\sum_{n \le N_\delta}\sum_{k=-n}^n \frac{ nr_e^{2n}|f_{n}^{k}|^2}{\rho^{2n}}\\
 & \ge \rho^{|n_{j(\alpha)+1}|-|n_{j(\alpha)}|}\sum_{k=-{n_{j(\alpha)}}}^{n_{j(\alpha)}}
 {n_{j(\alpha)}}r_*^{2{n_{j(\alpha)}}}|f_{n_{j(\alpha)}}^{k}|^2 \rightarrow \infty
 \end{align*}
 as $\alpha\to\infty$. So CALR takes place. This completes the proof of (i).

\medskip
To prove (ii) assume that $\Ge_c \neq -\Ge_s =1$. In this case, we have
$$
|b_n| \approx \frac{\rho^{2n}}{\delta + \rho^{2n}},\quad |c_n| \approx \frac{r_e^{2n}}{ \delta + \rho^{2n}},
$$
and
$$
E_\delta\approx\sum_{n=0}^\infty\sum_{k=-n}^n \frac{\Gd nr_e^{2n}|f_{n}^{k}|^2}{\delta^2+\rho^{4n}}.
$$
The rest of proof of (ii) is the same as that for (i).

\medskip
Suppose now that $-\Ge_s \neq 1$. If $\Ge_c=1$, then we have
$$
|b_n| \approx \frac{\rho^{2n}}{\delta + \rho^{2n}},\quad |c_n| \approx \frac{\Gd r_e^{2n}}{ \delta + \rho^{2n}},
$$
and
$$
E_\delta\approx\sum_{n=0}^\infty\sum_{k=-n}^n \frac{\Gd(\Gd^2+\rho^{2n}) nr_e^{2n}|f_{n}^{k}|^2}{(\delta+\rho^{2n})^2}\le\sum_{n=0}^\infty\sum_{k=-n}^n  nr_e^{2n}|f_{n}^{k}|^2 \le \left\| \frac{\p F}{\p r} \right\|_{L^2(\p B_e)}^2.
$$
Since the source function $f$ is supported outside the radius $r_e$, we have
$$
\left\| \frac{\p F}{\p r} \right\|_{L^2(\p B_e)} \le C \|f \|_{L^2(\Rbb^3)},
$$
and $E_\delta$ is bounded independently of $\Gd$. The case when $\Ge_c \neq 1$ can be treated similarly.

\appendix
\section{Gap property of dipoles}

In this appendix we show that the Newtonian potentials of dipole source functions satisfy the gap conditions [GC1] and [GC2]. We only prove [GC1] since the other can be proved similarly.

Let $f$ be a dipole in $B_{r_*} \setminus \overline{B}_e$, {\it i.e.},
$f(\Bx)=\Ba \cdot\nabla\delta_\By(\Bx)$ for a vector $\Ba$ and $\By\in B_{r_*}\setminus
\overline{B}_e$. Then its Newtonian potential is given by $F(\Bx)=-\Ba\cdot\nabla_{\By} G(\Bx-\By)$.
It is well-known (see, for example,
\cite{nedelec}) that the fundamental solution $G(\Bx-\By)$ admits the following expansion if $|\By| > |\Bx|$:
$$
G(\Bx-\By)= - \sum_{n=0}^\infty \sum_{k=-n}^{n} \frac{1}{2n+1}Y_n^k(\wBx) Y_n^k(\wBy) \,\frac{|\Bx|^n}{|\By|^{n+1}}.
$$
So we have
$$
F(\Bx)= \sum_{n=0}^\infty \sum_{k=-n}^{n} \frac{1}{2n+1}|\Bx|^n Y_n^k(\wBx) \, \Ba\cdot\nabla \left( \frac{1}{|\By|^{n+1}} Y_n^k(\wBy) \right),
$$
and hence
\beq\label{fnk}
f_n^k = \frac{1}{2n+1} \Ba\cdot\nabla \left( \frac{1}{|\By|^{n+1}} Y_n^k(\wBy) \right).
\eeq

We show that \beq \sum_{k=-n}^{n}  {n}r_*^{2{n}}|f_{n}^{k}|^2 \to
\infty \quad\mbox{as } n \to \infty, \eeq and hence [GC1] holds.
The following lemma is needed.

\begin{lem}
For any $\Ba$ and $\wBy$ on $S^2$ and for any positive integer $n$ there is a homogeneous harmonic polynomial $h$ of degree $n$ such that
\beq\label{dotone}
\Ba \cdot \nabla h(\wBy) = 1
\eeq
and
\beq\label{size}
\max_{|\wBx|=1} |h(\wBx)| \le \frac{\sqrt{3}}{n}.
\eeq
\end{lem}
\pf
After rotation if necessary, we may assume that $\wBy=(1,0,0)$. We introduce three homogeneous harmonic polynomials of degree $n$:
\begin{align*}
h_1(\Bx) &:= \frac{1}{2n} \left[ (x_1+ix_2)^n + (x_1-ix_2)^n \right], \\
h_2(\Bx) &:= \frac{1}{2ni} \left[ (x_1+ix_2)^n - (x_1-ix_2)^n \right], \\
h_3(\Bx) &:= \frac{1}{2ni} \left[ (x_1+ix_3)^n - (x_1-ix_3)^n \right].
\end{align*}
Then one can easily see that
$$
\nabla h_1(\wBy)=(1,0,0), \quad \nabla h_2(\wBy)=(0,1,0), \quad \nabla h_3(\wBy)=(0,0,1).
$$
So if we define
$$
h=a_1 h_1 + a_2 h_2 + a_3 h_3,
$$
then \eqnref{dotone} holds.

Since
$$
\max_{|\wBx|=1} |h_j(\wBx)| \le \frac{1}{n} \quad \mbox{for } j=1,2,3,
$$
we obtain \eqnref{size} using the Cauchy-Schwartz inequality. This completes the proof.
\qed

Let $\Ba$ and $\wBy$ be two unit vectors, and let $h$ be a homogeneous harmonic polynomial of degree $n$ satisfying \eqnref{dotone} and \eqnref{size}. Then $h$ can be expressed as
$$
h(\Bx)= \sum_{k=-n}^{n} \Ga_k |\Bx|^n Y_n^k(\wBx),
$$
where
\beq\label{ak}
\Ga_k = \frac{1}{4\pi} \int_{S^2} h(\wBx) Y_n^k(\wBx) dS.
\eeq
Because of \eqnref{dotone}, we have
$$
1 = \Ba \cdot \nabla h(\wBy) \le \sum_{k=-n}^{n} |\Ga_k| \left| \Ba \cdot \nabla \left( |\Bx|^n Y_n^k(\wBx) \right) \right|.
$$
So there is $k$, say $k_n$, between $-n$ and $n$ such that \beq
|\Ga_{k_n}| \left| \Ba \cdot \nabla \left( |\Bx|^n Y_n^{k_n}(\wBx)
\right) \right| \ge \frac{1}{2n+1}. \eeq On the other hand, from
\eqnref{size} and \eqnref{ak}, it follows by using Jensen's
inequality that
$$
|\Ga_{k_n}|^2 \le \frac{1}{4\pi} \int_{S^2} |h(\wBx)|^2 |Y_n^{k_n}(\wBx)|^2 dS  \le \frac{3}{n^2}.
$$
Thus we have
\beq
\left| \Ba \cdot \nabla \left( |\Bx|^n Y_n^{k_n}(\wBx) \right) \right| \ge \frac{n}{\sqrt{3}(2n+1)} \ge C
\eeq
for some $C$ independent of $n$.

Now one can see from \eqnref{fnk} that \beq |f_n^{k_n}| \ge
\frac{C}{n |\By|^{n+1}} \eeq for some $C$ independent of $n$.
Since $|\By| < r_*$, we obtain that
$$
\sum_{k=-n}^{n}  {n}r_*^{2{n}}|f_{n}^{k}|^2 \ge {n}r_*^{2{n}}|f_{n}^{k_n}|^2 \ge \frac{C}{n} \left( \frac{r_*}{|\By|} \right)^{2n} \to \infty \quad\mbox{as } n \to \infty,
$$
as desired. It is worth mentioning that the constants $C$ in the
above may be different at each occurrence, but are independent of
$n$.

\end{document}